# Determining the Fundamental Failure Modes in Ni-rich Lithium Ion Battery Cathodes

Siyang Wang[1,*], Zonghao Shen[1,2,3,†], Aigerim Omirkhan[1,2], Oriol Gavalda-Diaz[4,1], Mary P. Ryan[1,2], Finn Giuliani[1]

[1] Department of Materials, Royal School of Mines, Imperial College London, London, SW7 2AZ, UK

[2] The Faraday Institution, Quad One, Harwell Science and Innovation Campus, Didcot, OX11 0RA, UK

[3] Institute of Condensed Matter Chemistry of Bordeaux, (ICMCB)-CNRS, 87 Avenue du Docteur Schweitzer, 33608 Pessac, France

[4] Composites Research Group, University of Nottingham, Nottingham, NG7 2GX, UK

[*] Corresponding author: siyang.wang15@imperial.ac.uk

[†] Present address: Université Grenoble Alpes, CNRS, Grenoble INP, LMGP, 38000 Grenoble, France

## Abstract

Challenges associated with in-service mechanical degradation of Li-ion battery cathodes has prompted a transition from polycrystalline to single crystal cathode materials. Whilst for *single crystal* materials, dislocation-assisted crack formation is assumed to be the dominating failure mechanism throughout battery life, there is little direct information about their mechanical behaviour, and mechanistic understanding remains elusive. Here, we demonstrated, using *in situ* micromechanical testing, direct measurement of local mechanical properties within $LiNi_{0.8}Mn_{0.1}Co_{0.1}O_2$ single crystalline domains. We elucidated the dislocation slip systems, their critical stresses, and how slip facilitate cracking. We then compared single crystal and polycrystal deformation behaviour. Our findings answer two fundamental questions critical to understanding cathode degradation: What dislocation slip systems operate in Ni-rich cathode materials? And how does slip cause fracture? This knowledge unlocks our ability to develop tools for lifetime prediction and failure risk assessment, as well as in designing novel cathode materials with increased toughness in-service.

# Introduction

The successful development of next-generation Li-ion batteries (LIBs) exhibiting higher capacity relies on the stable functioning of cathode materials [1]. $LiNi_xMn_yCo_{1-x-y}O_2$ (NMC) is a favourable material family owing to their high capacity and operating voltage, yet relatively low cost [2–4]. $LiNi_{0.8}Mn_{0.1}Co_{0.1}O_2$, or NMC811, stands out due to its high specific energy density and low Co content [5]. The biggest challenge is the rapid in-service degradation of Ni-rich NMC particles, leading to significantly reduced capacity over time [6–8]. Apart from the chemical degradation processes, observation of intergranular fracture events in agglomerated polycrystalline particles post-cycling has prompted the recent development of single crystal particles in LIB cathodes [9,10]. This has moderately improved mechanical stability [11] and efficiency of Li transport, however the evidence of intragranular defect generation suggests that durability is still a challenge [12].

The mechanical properties of NMC particles contribute significantly to their chemical and electrochemical behaviour, thereby affecting the capacity and cyclability of LIBs. It is believed that mechanical degradation is an undoubted reason for battery failure [7]. Mechanical degradation, or more specifically, cracking, increases the total surface area of the particles. The fresh and free surface will convert into rock-salt layer from layered structure with increased oxygen loss. Studies have proved that the released oxygen is singlet oxygen, which is very active and can react with the electrolyte to generate moisture and HF [13,14]. HF can attack the active cathode material, in this case NMC811, which expedites transition metal dissolution and further oxygen loss. Moreover, formation of the rock-salt layer increases the ionic and electric impedance of the cathode, leading to sluggish Li-ion diffusion. Thus, mechanical failure of NMC ultimately results in degraded capacity and cyclability of LIBs.

Two main sources of mechanical degradation of these materials have been reported:

1. NMC particles are bonded to the Al current collector *via* tape casting followed by a calendaring process, to improve particle packing density and to enhance the particle-current collector connection. The mechanical load experienced during calendaring can however cause particle cracking which subsequently deteriorates battery performance [15,16].

2. Charge and discharge cycles involve the removal and insertion of Li ions in the basal planes ("Li slabs") of the crystal lattice (Figure 1(a)). This changes lattice parameter and therefore shape of the crystals [5]. For polycrystalline particles, this may cause intergranular fracture [17]. For single crystal particles, this, along with concentration gradient of Li atoms, may trigger microcrack formation and dislocation movement [12].

There is however controversy over whether the Li-intercalation-induced stress is significant enough to cause defect formation and mobility. The strength of $LiCoO_2$ single crystals (with the same crystal structure as NMC) has been measured at 2-4 GPa [18], but shear stresses involved upon charge/discharge are an order of magnitude lower according to numerical simulations based on an isotropic diffusion-induced stress model [12]. Stallard *et al.* who employed an indirect mechanical testing approach has argued that the basal plane shear strength of NMC811 is below 100 MPa [19]. This is not supported by the results of Feng *et al.* [18] where none of the samples, even the cycled ones, showed sub-GPa-level strength. Therefore the basal plane shear strength is unlikely to be much lower than 1 GPa, unless the orientations of all samples tested, although not reported, coincidentally had their basal planes either parallel or perpendicular to the loading axis, making Schmid factor for basal slip zero [20].

Numerous research articles report the detrimental effect of dislocations in cathode materials on the electrochemical performance of batteries [21–27]. In crystalline materials, dislocations are a group of 1D defects, and their movement at the atomic scale leads to plastic deformation at a larger scale. Under load, dislocations often glide on specific sets of crystallographic planes. These planes, and the dislocation Burgers vectors, are often material-dependent and are termed slip systems. Slip systems, and the stresses required to activate them, fundamentally determine the mechanical performance of a (plasticity-controlled) material [28] such as NMC811. In LIB cathodes, dislocations form due to electrochemically and/or thermally induced strain fields upon synthesis and cycling [23,27], and they are found to facilitate crack formation through two mechanisms:

1. Cracks are formed at edge dislocation cores directly due to the local stress fields [22–25].

2. Dislocation glide results in oxygen release [21] which routinely triggers crack formation in battery materials [25].

However, work in the literature to date has failed to answer key questions about how dislocations facilitate cracking of NMC materials quantitatively: What are the slip systems, failure modes, and just how strong are each of these modes? Also, where is the critical point during dislocation glide that cracks start to form? Answers to these questions can form the basis of understanding mechanical behaviour of any material [28] and potentially expedite the improvement of materials and device design against degradation [29]. They are also key to understanding failure at *all* stages of battery life, as the (dislocation-assisted crack formation) failure mechanism operates independently of the stress state and the source of stress.

In LIBs, NMC materials are often in the form of powders [10]. Mechanical testing of powders, by compressing them with a flat indenter, can be used to evaluate performance [30]. However, converting load-displacement data into stress-strain curves can result in large uncertainties given the irregular shapes of the samples, and extraction of deformation modes is difficult as the crystal orientations are largely unknown. Fabrication of small-scale samples with well-defined geometries and known crystal orientations is hence essential, and current state-of-the-art equipment allows testing to be carried out *in situ* inside electron microscopes [31–33] where failure can be imaged directly in real-time. Here, we carried out *in situ* compression tests of pristine NMC811 single crystal micropillars with known crystal orientations in a scanning electron microscope (SEM), to calculate the slip systems in this material, their critical stresses, and how dislocations gliding on different slip systems trigger crack formation.

NMC811 powders were sintered into a bulk material to produce a stiff substrate for mechanical testing. X-ray diffraction (XRD) was used to identify the crystal structure (Figure 1(b)) and to measure lattice parameters (Table S1). Transmission electron microscopy (TEM) was used to observe the layered atomic arrangement (Figure 1(c)), and secondary ion mass spectrometry (SIMS) to probe the variations in chemical composition before/after thermal treatment (Figure S1). Electron backscatter diffraction (EBSD) analysis was conducted to reveal the microstructure and crystal orientations (Figure 1(d)), for fabricating and testing micropillars within targeted crystals. Pillars with dimensions on the same length scale as the single crystal particles employed in real batteries were fabricated and tested in order to rule out any size effects in micromechanical testing [34]. Thereafter, the slip systems were identified based on knowledge of the crystal orientations, and the stress required to activate each of the slip systems was calculated. This work enabled detailed analysis of how deformation gives rise to failure, and allowed compared behaviours between single crystal and polycrystalline samples.

# Methods

## Sample preparation

Sintered pellets were prepared with commercially available single crystal NMC811 powders from Li-Fun Technology Co. (Figure S2(a,b)), which were stored in an Ar-filled glovebox (< 0.6 ppm $H_2O$; < 0.6 ppm $O_2$). There are two reasons why the powders were sintered into a bulk pellet before the mechanical tests were carried out:

1. It is not practically feasible to polish the powders for EBSD experiments, making it difficult to determine the crystal orientations pre-test, and therefore activated slip systems of test pieces post-test.

2. A sintered pellet can serve as a stiff substrate for the micromechanical tests, so as to substantially reduce the compliance of the test setup (compared to stacked powders).

The powders were firstly ground and mixed with $Li_2CO_3$ (10% Li excess) in the mortar and then uniaxially pressed into pellets in the glovebox. The pellets were further isostatically pressed at 300 MPa before sintering in a Pt crucible at 1000 °C for 10 h in static air [35]. The sintering temperature is higher than the decomposition temperatures for potential surface impurities such as $Li_2CO_3$ and LiOH which are detrimental to capacity retention [36,37]. To remove any residual Li impurities on the surface after sintering, the sintered pellets were ground using SiC papers with successive grades in the glovebox and cleaned in ethanol ultrasonic bath before further characterisation.

## XRD

XRD measurements were performed with a D2 Phaser Diffractometer (Cu Kα) between 10° and 80° at a step size of 0.034°. Lattice parameters were obtained *via* Le Bail refinement with the Fullprof suite [38].

## TEM

Thin foil specimens were prepared using focussed ion beam (FIB) *in situ* lift-out technique, on a Thermo Fisher Scientific (TFS) Helios 5 CX DualBeam microscope. TEM characterisation was performed on a JEOL JEM-2100F field emission gun source transmission electron microscope at an acceleration voltage of 200 kV.

## SIMS

An ION TOF TOF-SIMS[5] time-of-flight secondary ion mass spectrometer was employed. Analysis was performed in the negative mode with the high current bunch mode at the mass resolution of *ca.* 10000. A 25 keV $Bi^+$ primary beam was used for the analysis over an area of 100 × 100 $\mu m^2$ and a 500 eV single $Ar^+$ sputtering beam was applied for sputtering over an area of 300 × 300 $\mu m^2$.

## EBSD

We employed EBSD to determine the crystal orientations of the grains in the sintered pellet. We mapped a large area on the polished sample surface, and then looked for grains with desired orientations for micromechanical testing. The sample frame (x,y,z directions) was physically marked and kept constant throughout the EBSD – FIB milling – *in situ* testing – post-mortem analysis procedure, which allowed us to 1) fabricate and test pillars in grains with desired orientations and 2) work out the crystallographic planes and directions (slip systems) associated with the slip traces.

A sintered pellet was polished mechanically with diamond suspension (in ethanol), and then polished with broad ion beam in a Gatan PECS II Ar ion polishing system. EBSD characterisation was carried out on a Zeiss Sigma 300 SEM equipped with a EDAX Clarity™ Plus direct electron detector EBSD camera, using a beam acceleration voltage of 20 kV, a probe current of ~10 nA, and a spatial step size of 0.25 μm. EBSD data was analysed using EDAX OIM Analysis v8 software. The raw data (IPF map in Figure S3) was processed through dictionary indexing of the EBSD patterns [39] to help resolve pseudo symmetry of the crystal structure, followed by a clean-up step with a single iteration of grain dilation.

## *In situ* micropillar compression tests

Micropillars were fabricated with FIB milling on the Helios 5 CX DualBeam microscope, automated via employing TFS NanoBuilder Software (Figure S2(c)). The single crystal pillars are ~2.5 μm in height and 1 μm in mid-height diameter, while the polycrystalline pillars are ~15 μm in height and 5 μm in mid-height diameter. The taper angle is ~5°. For the 30 keV $Ga^+$ FIB used to mill the pillars, the FIB-damaged zone with Ga-ion implantation varies with material, but is typically on the order of ~10 nm [40–46], about 1% of the pillar thickness. Thus the effect of the damaged zone on the mechanical responses of the pillars should be nearly negligible.

*In situ* compression tests of the micropillars were conducted using an intrinsically displacement-controlled Alemnis nanoindenter in a TFS Quanta 650 SEM. A schematic diagram of the experimental setup is shown in Figure S2(d). The single crystal and polycrystalline pillars were tested at displacement speeds of 5 nm/s and 30 nm/s respectively, resulting in the same strain rates for both sets of tests. The total displacement applied for each pillar can be found on the stress-displacement curves. No hold at maximum displacement was applied (loading was followed immediately by unloading). Stress is defined as load divided by the top surface cross-sectional area of each pillar. When the tests were complete, post-deformation SEM images of the micropillars were captured on a Zeiss Sigma 300 SEM, using an acceleration voltage of 5 kV and the in-lens detector for achieving high spatial resolution. The slip system(s) activated for each pillar were then identified based on the orientation(s) of the slip bands and the crystal orientation derived from EBSD data.

It is worth noting that the displacement rates were selected such that the tests can be completed in minutes. This is important for small scale mechanical testing where thermal drift of load/displacement sensors may significantly affect the test results, especially when working with low load levels. Long experiments may invalidate the linear load drift assumption one normally makes when correcting raw data. Hence the displacement rate used could be higher than that often experienced in actual battery cathodes upon cycling. For example, if during delithiation of an NMC particle, a contraction of 10 % and a charge rate of 1 C are assumed, this would result in a deformation rate of ~7e-2 nm/s which is smaller than the 5 nm/s that were used in the manuscript. Strain rate sensitivity of the failure process could be studied in detail in future work.

# Results and discussion

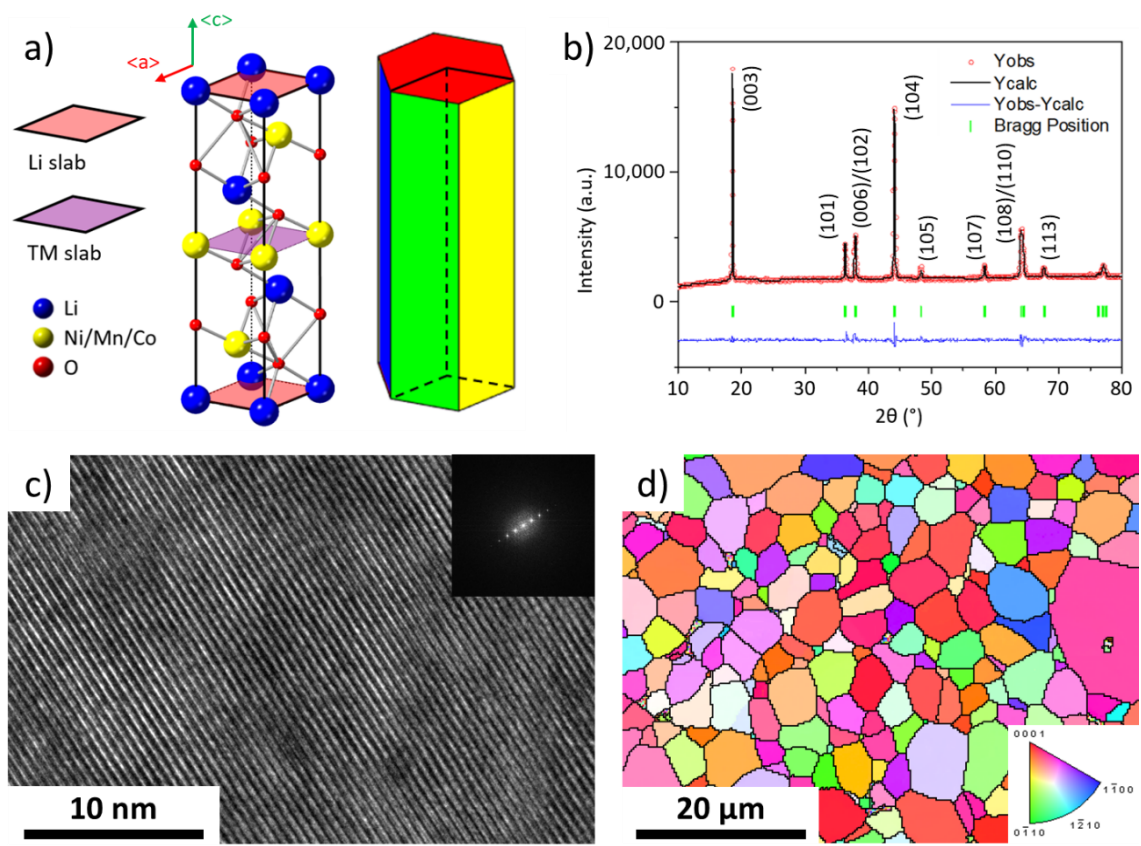

Figure 1 XRD and electron microscopy characterisation of the material studied. (a) Crystal structure of $LiNi_xMn_yCo_{1-x-y}O_2$ (space group R-3m), with spheres showing the positions of the atoms and shaded planes showing the Li and transition metal (TM) slabs. The <a> and <c> directions are labelled. For simplicity, hexagons (on the right) are adopted in this work to represent crystal orientations of micropillars. (b) XRD pattern and refinement of the sintered NMC811 pellet revealing crystal structure retention after sintering. (c) Bright field TEM image of NMC811 showing the layered structure at atomic scale, with an insert showing the fast Fourier transformation (FFT). (d) EBSD inverse pole figure (IPF)-*z* map (*z* points out of the page / sample surface) of sintered NMC811 showing grain structure and crystal orientations.

To activate basal slip, a single crystal pillar with its basal plane ~45° to the loading axis was deformed. An SEM image of the deformed (to a displacement of ~220 nm) pillar is shown in Figure 2(a), which indicates that plastic slip occurred on the basal plane and along the <a> direction. A long vertical crack is observed beneath a slip band. The crack is through-thickness as evidenced by Figure S4(a) where the same pillar is viewed from another angle. The crack appears to be non-straight and is therefore unlikely along a specific crystallographic plane.

The pillar in Figure 2(b) which has its basal plane normal (*c*-axis) nearly perpendicular (~85°) to the loading axis, showed similar behaviour: plastic slip on an inclined crystallographic plane and a vertical crack beneath the slip plane. In contrast, plastic slip on this pillar occurred on the prismatic plane and along the <a> direction. Additionally, the vertical crack was evidently connected to a shear crack on the plane of plastic slip, same as the crack on the previous pillar (Figure S4(a)).

Figure 2(c) shows a pillar with its c-axis close (~17°) to the loading direction. Unlike the other samples, this pillar exhibited only plastic slip without cracking. For this pillar, slip occurred on a pyramidal plane (see the shaded plane on the hexagon in Figure 2(c) which is an inclined plane between the Li and TM slabs, Figure 1(a)) and the slip direction is <½c + a>. Figure S4(b) is an image of the same pillar viewed from another angle, which also shows no evidence of cracking. However, a secondary slip system, also of <½c + a> pyramidal type, can be observed and is marked in Figure S4(b$_2$).

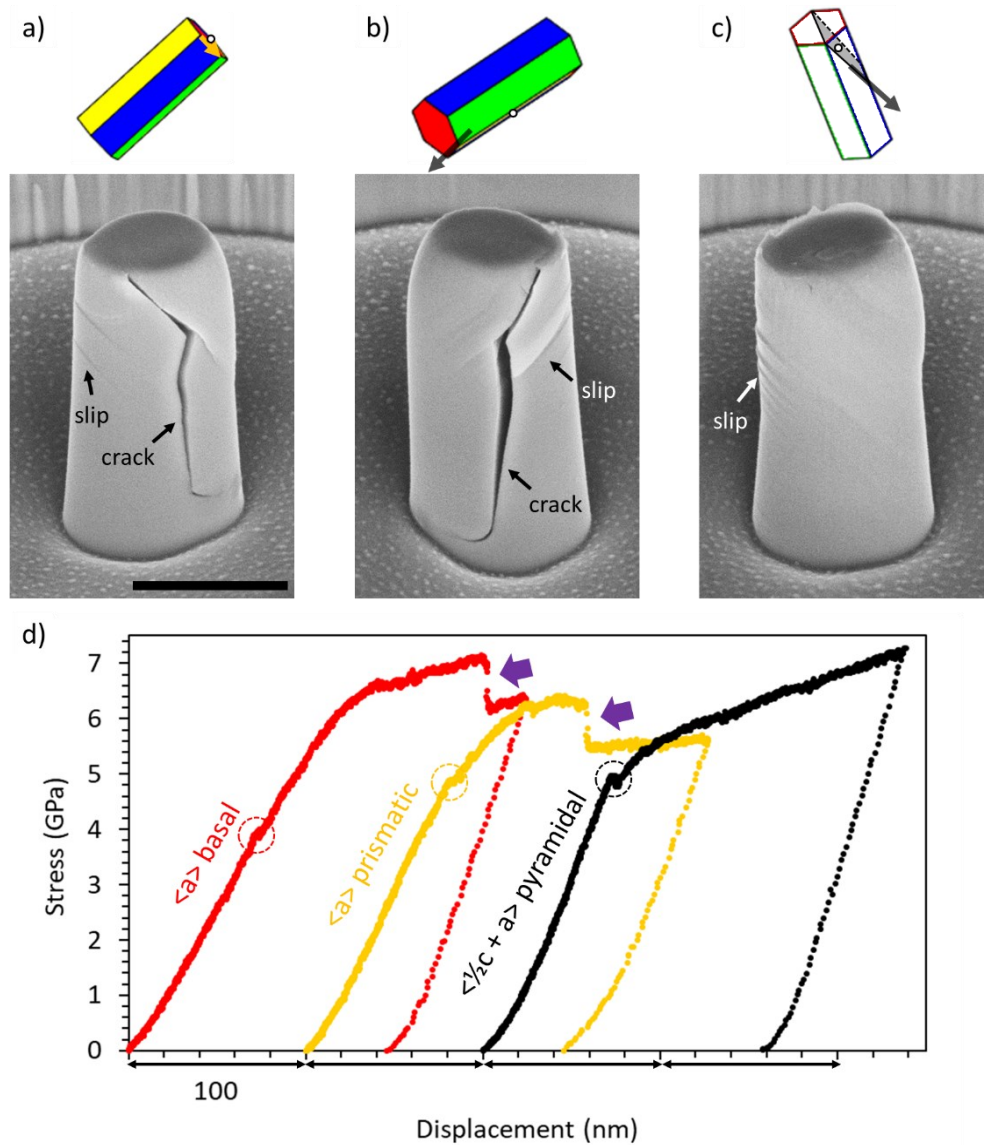

Figure 2 Post-deformation SEM images and mechanical responses of single crystal NMC811 micropillars showing activation of single (types of) slip systems. Images of micropillars with basal plane normal (a) ~45°, (b) ~perpendicular (~85°) and (c) close (~17°) to the loading axis, with hexagons showing the crystal orientations (see Figure 1(a)). On each of the hexagons, the dot and the arrow indicate the slip plane and direction, respectively. Both plastic slip and cracking events can be observed on pillars in (a) and (b). The pillar in (c) shows plastic slip without apparent cracking. The scale bar is 1 μm and the same for all SEM images. (d) Stress-displacement curves recorded during compression tests of the micropillars. The red, yellow and black curves are for the pillars in (a), (b), and (c), respectively. Stress drops can be observed on the red and the yellow curves, as marked by the purple arrows, and correspond to the initiation of pillar cracking evidenced in (a) and (b).

The mechanical responses of the above pillars are plotted in Figure 2(d). Some distinct features are observed:

1. Each of the pillars experienced a noticeable amount of stable plastic deformation (> 70 nm) prior to any stress drop.
2. The yield stress of the pillar exhibiting basal slip is only gently lower than the other pillars, meaning that basal slip is not significantly weaker than the other deformation modes, contrary to the hypothesis made elsewhere [19] and in contrast to the behaviours of some other layered materials [47,48].
3. Stress drop events, as marked by the purple arrows, are present on the curves for the two pillars exhibiting basal and prismatic slip respectively, corresponding to the process of crack development as observed *in situ* (Figure 2(a,b)).

Table 1 The CRSS for <a> basal, <a> prismatic, and <½c + a> pyramidal slip systems in NMC 811, calculated through multiplying the yield stress (the stress at the first observed deviation from linear elastic region, as marked by the dashed circles in Figure 2(d)) of each of the pillars in Figure 2 by the Schmid factor for the slip system activated.

|  | <a> basal | <a> prismatic | <½c + a> pyramidal |
| --- | --- | --- | --- |
| Yield stress (GPa) | 4.01 | 4.97 | 4.95 |
| Schmid factor | 0.42 | 0.47 | 0.48 |
| CRSS (GPa) | 1.7 | 2.3 | 2.4 |

The critical resolved shear stresses (CRSS) for the three types of slip systems activated was extracted from the yield stresses and the crystal orientations of the pillars using Schmid's law (Table 1). Despite the highly low-symmetry crystal structure (Figure 1(a)), the shear stresses required to activate the basal, prismatic and pyramidal slip systems are relatively close (1.7, 2.3 and 2.4 GPa respectively), giving rise to the only moderately anisotropic yield strength levels (Figure 2(d)).

Nevertheless, the post-yield behaviour was fairly *anisotropic*: the orientations that provoke single slip (Figure 2(a) and (b) where only one basal and one prismatic slip system was activated respectively) were seen to trigger crack formation that tends to cause complete fracture upon further loading. Due to the incompressible nature of plasticity [49], horizontal strain components were generated under uniaxial stress to retain the volumes of the pillars upon plastic deformation, which could drive the formation of these vertical cracks in brittle materials [50,51]. Such cracks were also observed in compression tests of agglomerated polycrystalline NMC811 particles (Figure S5). Because of the instant nature of the crack formation process (see the load drops in Figure 2(d)), it was not possible to determine if the vertical or the shear crack occurred first *in situ* (frame time was ~1.6 s). A possibility is that the crack orientation was firstly perpendicular to the slip plane due to a widening of the atom layers on one side of the slip plane, and afterwards changes its orientation to a vertical orientation because of indirectly resulting horizontal tensile strain. Nonetheless confirming this, requires a more stable crack growth process and higher resolution real-time characterisation. This may be achievable through *in situ* fracture tests in the TEM which were recently demonstrated on other materials by some of the authors of this work [52]. No long vertical crack was observed on the pillar in Figure 2(c)/Figure S4(b) where more than one pyramidal slip system operated. Multiple slip in this case is likely due to the (nearly) equal Schmid factors for all pyramidal slip systems as a result of the crystal orientation.

In addition, cooperation of multiple slip systems in single tests were observed (Figure 3). The coactivation of basal, prismatic and pyramidal slip systems can be observed on the pillar in Figure 3(a). Similar to the pillar in Figure 2(c) which also involved multiple slip, no evidence of cracking was found on this pillar.

Figure 3(b) presents another pillar exhibiting multiple slip. Unlike the sample in Figure 3(a), this pillar showed two vertical cracks post-deformation, at intersections between slip bands. These cracks, however, appear to be significantly shorter than those seen in Figure 2(a) and (b). They are also not through-thickness as evidenced by their absence in Figure S4(c) which shows the rear side of the same pillar. Another example is shown in Figure S4(d), where a short crack was observed at intersections between two sets of slip bands on pyramidal planes (the orientation of this pillar is similar to that in Figure 2(c) where the c-axis is ~parallel (~10°) to the sample surface). These data therefore suggest that single slip might trigger the generation of long and through-thickness cracks that strongly affect structural integrity of the crystals (Figure 2(a,b)), whereas multiple slip could suppress the formation of such large cracks (Figure 2(c), Figure 3(a)) although much smaller cracks can form at slip band intersections (Figure 3(b), Figure S4(d)). These smaller cracks do not immediately lead to failure, yet they may harm local electron transport upon battery operation, interact with electronic defects and develop into larger defects during charge/discharge cycles causing safety issues.

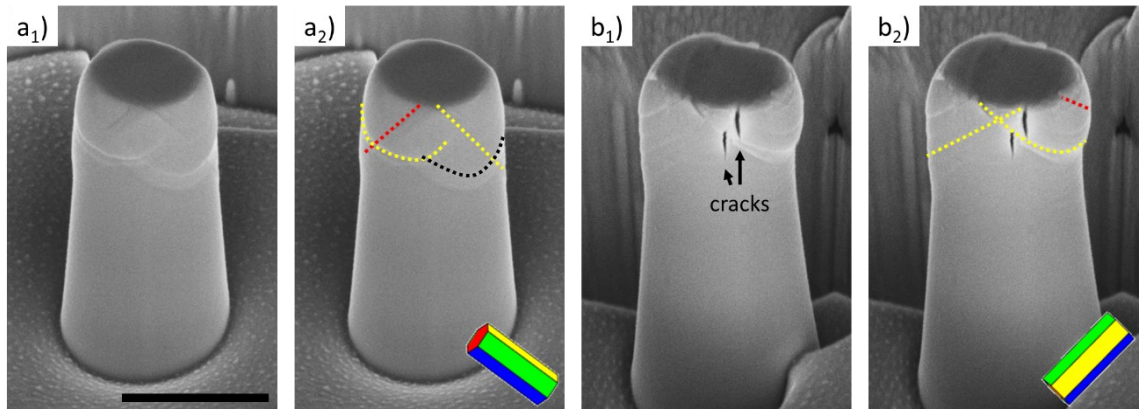

Figure 3 Post-deformation SEM images of single crystal NMC811 micropillars showing coactivation of multiple slip systems. Slip traces on the pillars in ($a_1$) and ($b_1$) are marked with dotted lines in ($a_2$) and ($b_2$), respectively. Basal, prismatic and pyramidal slip traces are marked in red, yellow and black, respectively. Two vertical cracks can be observed on the pillar in (b), at intersections between slip bands. The scale bar is 1 μm for all images.

Our results indicate moderately anisotropic yield strength of single crystal NMC811, where basal slip is only gently weaker than other deformation modes. This explains prior work where pristine $LiCoO_2$ single crystals with (presumably) random orientations always exhibit a similar strength level to those in Figure 2(d) [18]. However, when subjected to Li-intercalation-induced stresses of a few hundred MPa [12], such a strong material should not obviously degrade. This suggests that damage accumulation upon cycling is possibly accelerated by pre-existing defects in the material and/or (electro-)chemical driving forces which can induce various types of defects [53]. During calendaring where the particles are subjected to compression and shear, defects similar to those observed on the pillars may be generated, which could decrease the threshold stress required for fracture, and therefore the Li-intercalation-induced stresses could indeed be high enough to provoke failure. During electrochemical cycling, chemical degradation processes, *e.g.*, oxygen-loss-induced phase transformation, transition metal dissolution *etc.*, can change the materials chemistry (and phase structure) and further affect the mechanical properties. Future work on *in situ* and *operando* characterisation of uncalendared/calendared particles during charge cycles could help clarify the problem.

During electrochemical cycling, the stresses in the cathode particles are sometimes tensile. One may argue that this might lead to a different mechanical response to the results of the compression tests here in ceramic-like materials such as NMC, and that under tensile stresses, the material may not fail due to plastic deformation caused by shear stresses, but due to brittle failure caused by normal stresses. However, many prior works reported dislocation-assisted crack formation in LIB cathodes after cycling [21–25], indicating the dominance of this mechanism regardless of the stress state. Therefore the slip systems, critical stresses and fracture modes extracted in this work should be useful for understanding/modelling cathode failure not only upon calendaring where the stress state is compressive, but also upon cycling where tensile stresses occur.

Polycrystalline pillars were also tested to understand the role of grain boundaries (GB) in the failure of sintered NMC811. Note that sintered NMC studied here is different to common agglomerated polycrystalline NMC [10], and the latter should be much weaker at GBs as no "diffusion bonding" treatment is applied. As Figure 4 shows, generally the displacement applied was accommodated by fracture of GBs as opposed to slip inside grains, and the fracture only happens in a brittle way along the GBs. Local fracture mode is dependent on the GB orientation with respect to external load: shear fracture along GBs are observed for those inclined to the loading axis (Figure 4(a)), while GBs parallel to the loading axis tend to split under compression (Figure 4(b)). Plastic slip was observed in one case where it occurred shortly after an adjacent GB shear fracture event (Figure 4(c)). Although the strength level of the polycrystalline pillars is typically over 1 GPa for the dimensions employed, collapse of the pillars occurred quickly after GB fracture for all cases. Thus, sintered polycrystalline NMC811, although reasonably strong, exhibited limited ductility compared to single crystal NMC811 as also evidenced by the significant stress drops on the stress-displacement curves (Figure S7).

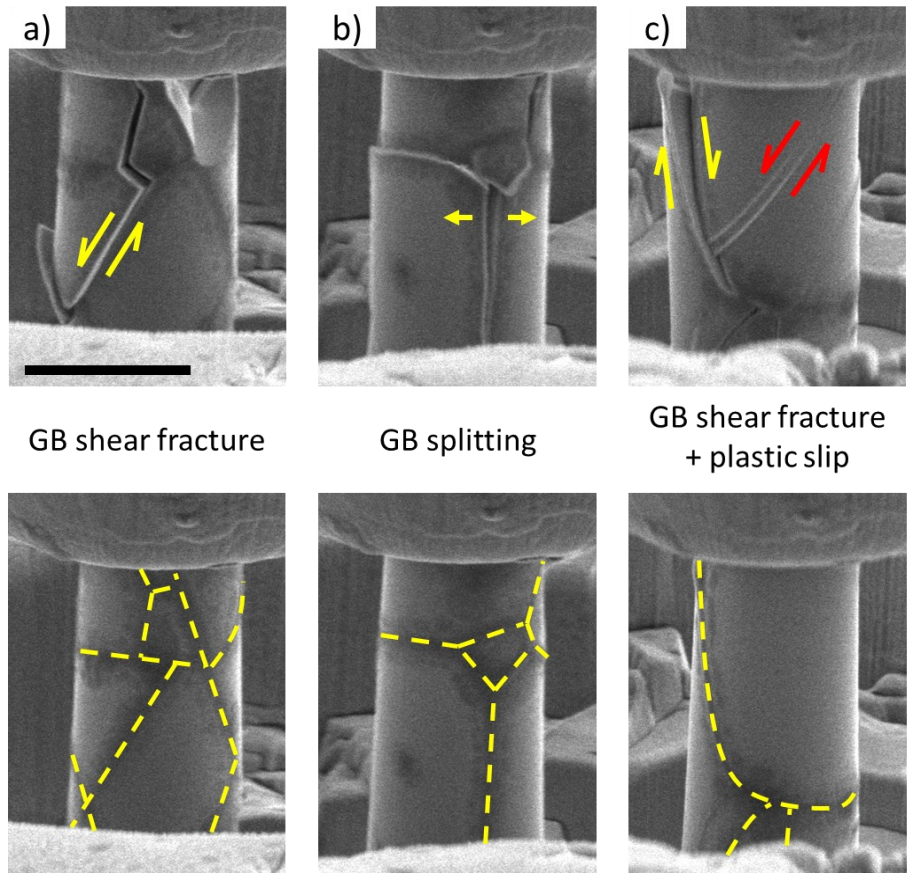

Figure 4 The deformation behaviours of sintered polycrystalline NMC811 micropillars. Images showing (a) shear fracture along GBs inclined to the loading axis, (b) splitting of GBs parallel to the loading axis and (c) both shear fracture along GBs and plastic slip inside a grain. Deformation of grain boundaries is marked with yellow arrows, and plastic slip is marked with red arrows. Images of the pillars before testing are shown to highlight the GBs (with the yellow dashed lines). The scale bar is 5 μm for all images.

## Conclusions

In order to gain quantitative knowledge of the fundamental deformation and failure modes of pristine NMC811 as a cathode material for next-generation LIBs, we carried out *in situ* SEM micromechanical tests on samples with known crystal orientations and a well-defined stress state. The following conclusions can be drawn:

1. At room temperature, μm-scale NMC811 single crystals exhibit dislocation glide under uniaxial compression, which can provoke fracture upon further loading.
2. Three types of slip systems were observed: <a> basal, <a> prismatic and <½c + a> pyramidal.

3. Despite the highly low-symmetry crystal structure, the CRSS for basal slip (1.7 GPa) is only slightly lower than those for prismatic (2.3 GPa) and pyramidal (2.4 GPa) slip systems, making the yield strength of single crystal NMC811 moderately anisotropic.

4. Under uniaxial compression, long cracks are often observed upon activation of single slip systems, while coactivation of multiple slip systems might give rise to more stable plastic flow and higher ductility, though small cracks can form at slip band intersections.

5. Therefore, particles with c-axis pointing out-of-plane could be less susceptible to significant cracking upon calendaring, due to the high likelihood of triggering co-activation of multiple pyramidal slip systems.

6. Stresses induced upon charge cycles are *unlikely* the sole origin of mechanical degradation of single crystal NMC811, but may accelerate failure if pre-existing defects are generated during cathode manufacturing or other mechanical impacts on the system.

7. Deformation of sintered polycrystalline NMC811 is mostly accommodated by brittle fracture of grain boundaries with minimal plastic slip, and therefore showed negligible ductility whilst exhibiting GPa-level strength.

## Data availability

The data that support the findings of this study are available from the corresponding author upon reasonable request.

## Acknowledgements

ZS, AO and MPR acknowledge funding from The Faraday Institution (G116198MBAG/740). SW and FG acknowledge funding from Engineering and Physical Science Research Council (EPSRC) through MAPP project (EP/P006566/1). MPR acknowledges funding from the Armourers and Brasiers' Company. The FIB used was funded by EPSRC as part of the cryo-EPS facility at Imperial College London (EP/V007661/1). The TFS Quanta SEM used was supported by the Shell AIMS UTC and is housed in the Harvey Flower EM suite at Imperial College London. We thank Dr Katharina Marquardt, Dr Matt Nowell, Dr William Lenthe, Prof Ben Britton and Dr Ruth Birch for discussions. The Matlab code used to plot unit cells for representation of crystal orientations was written by Dr Vivian Tong at National Physical Laboratory UK.

# Author contributions

SW and FG designed the study. SW performed EBSD analysis and mechanical testing, and wrote the original draft. ZS prepared the sample and carried out XRD and SIMS experiments. AO and OGD conducted TEM imaging. MPR and FG supervised the work. All authors reviewed the manuscript.

# Competing interest declaration

The authors declare no competing interests.

# References

[1]   L.S. De Vasconcelos, R. Xu, Z. Xu, J. Zhang, N. Sharma, S.R. Shah, J. Han, X. He, X. Wu, H. Sun, S. Hu, M. Perrin, X. Wang, Y. Liu, F. Lin, Y. Cui, K. Zhao, Chemomechanics of Rechargeable Batteries: Status, Theories, and Perspectives, Chem. Rev. 122 (2022) 13043–13107. https://doi.org/10.1021/acs.chemrev.2c00002.

[2]   J.B. Goodenough, K.S. Park, The Li-ion rechargeable battery: A perspective, J. Am. Chem. Soc. 135 (2013) 1167–1176. https://doi.org/10.1021/ja3091438.

[3]   S.T. Myung, F. Maglia, K.J. Park, C.S. Yoon, P. Lamp, S.J. Kim, Y.K. Sun, Nickel-Rich Layered Cathode Materials for Automotive Lithium-Ion Batteries: Achievements and Perspectives, ACS Energy Lett. 2 (2017) 196–223. https://doi.org/10.1021/acsenergylett.6b00594.

[4]   A. Manthiram, A reflection on lithium-ion battery cathode chemistry, Nat. Commun. 11 (2020) 1–9. https://doi.org/10.1038/s41467-020-15355-0.

[5]   K. Märker, P.J. Reeves, C. Xu, K.J. Griffith, C.P. Grey, Evolution of Structure and Lithium Dynamics in LiNi0.8Mn0.1Co0.1O2 (NMC811) Cathodes during

Electrochemical Cycling, Chem. Mater. 31 (2019) 2545–2554. https://doi.org/10.1021/acs.chemmater.9b00140.

[6]  Y. Dong, J. Li, Oxide Cathodes: Functions, Instabilities, Self Healing, and Degradation Mitigations, Chem. Rev. (2022). https://doi.org/10.1021/acs.chemrev.2c00251.

[7]  R. Xu, H. Sun, L.S. de Vasconcelos, K. Zhao, Mechanical and Structural Degradation of LiNixMnyCozO2 Cathode in Li-Ion Batteries: An Experimental Study, J. Electrochem. Soc. 164 (2017) A3333–A3341. https://doi.org/10.1149/2.1751713jes.

[8]  L.S. de Vasconcelos, N. Sharma, R. Xu, K. Zhao, In-Situ Nanoindentation Measurement of Local Mechanical Behavior of a Li-Ion Battery Cathode in Liquid Electrolyte, Exp. Mech. 59 (2019) 337–347. https://doi.org/10.1007/s11340-018-00451-6.

[9]  J. Li, H. Li, W. Stone, R. Weber, S. Hy, J.R. Dahn, Synthesis of Single Crystal LiNi0.5Mn0.3Co0.2O2 for Lithium Ion Batteries, J. Electrochem. Soc. 164 (2017) A3529–A3537. https://doi.org/10.1149/2.0401714jes.

[10] J. Li, A.R. Cameron, H. Li, S. Glazier, D. Xiong, M. Chatzidakis, J. Allen, G.A. Botton, J.R. Dahn, Comparison of Single Crystal and Polycrystalline LiNi0.5Mn0.3Co0.2O2 Positive Electrode Materials for High Voltage Li-Ion Cells, J. Electrochem. Soc. 164 (2017) A1534–A1544. https://doi.org/10.1149/2.0991707jes.

[11] J.E. Harlow, X. Ma, J. Li, E. Logan, Y. Liu, N. Zhang, L. Ma, S.L. Glazier, M.M.E. Cormier, M. Genovese, S. Buteau, A. Cameron, J.E. Stark, J.R. Dahn, A Wide Range of Testing Results on an Excellent Lithium-Ion Cell Chemistry to be used as Benchmarks for New Battery Technologies, J. Electrochem. Soc. 166 (2019) A3031–A3044. https://doi.org/10.1149/2.0981913jes.

[12] Y. Bi, J. Tao, Y. Wu, L. Li, Y. Xu, E. Hu, B. Wu, J. Hu, C. Wang, J.G. Zhang, Y. Qi, J. Xiao, Reversible planar gliding and microcracking in a single-crystalline Ni-rich cathode, Science (80-. ). 370 (2020) 1313–1318. https://doi.org/10.1126/science.abc3167.

[13] J. Wandt, A.T.S. Freiberg, A. Ogrodnik, H.A. Gasteiger, Singlet oxygen evolution from layered transition metal oxide cathode materials and its implications for lithium-ion batteries, Mater. Today. 21 (2018) 825–833. https://doi.org/10.1016/j.mattod.2018.03.037.

[14]  R. Jung, M. Metzger, F. Maglia, C. Stinner, H.A. Gasteiger, Oxygen Release and Its Effect on the Cycling Stability of LiNixMnyCozO2 (NMC) Cathode Materials for Li-Ion Batteries, J. Electrochem. Soc. 164 (2017) A1361–A1377. https://doi.org/10.1149/2.0021707jes.

[15]  T.M.M. Heenan, A. Wade, C. Tan, J.E. Parker, D. Matras, A.S. Leach, J.B. Robinson, A. Llewellyn, A. Dimitrijevic, R. Jervis, P.D. Quinn, D.J.L. Brett, P.R. Shearing, Identifying the Origins of Microstructural Defects Such as Cracking within Ni-Rich NMC811 Cathode Particles for Lithium-Ion Batteries, Adv. Energy Mater. 10 (2020). https://doi.org/10.1002/aenm.202002655.

[16]  C. Meyer, M. Kosfeld, W. Haselrieder, A. Kwade, Process modeling of the electrode calendering of lithium-ion batteries regarding variation of cathode active materials and mass loadings, J. Energy Storage. 18 (2018) 371–379. https://doi.org/10.1016/j.est.2018.05.018.

[17]  D.J. Miller, C. Proff, J.G. Wen, D.P. Abraham, J. Bareño, Observation of microstructural evolution in li battery cathode oxide particles by in situ electron microscopy, Adv. Energy Mater. 3 (2013) 1098–1103. https://doi.org/10.1002/aenm.201300015.

[18]  L. Feng, X. Lu, T. Zhao, S. Dillon, The effect of electrochemical cycling on the strength of LiCoO2, J. Am. Ceram. Soc. 102 (2019) 372–381. https://doi.org/10.1111/jace.15893.

[19]  J.C. Stallard, S. Vema, D.S. Hall, A.R. Dennis, M.E. Penrod, C.P. Grey, V.S. Deshpande, N.A. Fleck, Effect of Lithiation upon the Shear Strength of NMC811 Single Crystals, J. Electrochem. Soc. 169 (2022) 040511. https://doi.org/10.1149/1945-7111/ac6244.

[20]  E. Schmid, W. Boas, Plasticity of Crystals With Special Reference to Metals, 1st ed., F. A. Hughes & Co. Limited, London, 1950. https://doi.org/10.1107/S0567739469001306.

[21]  Q. Li, Z. Yao, E. Lee, Y. Xu, M.M. Thackeray, C. Wolverton, V.P. Dravid, J. Wu, Dynamic imaging of crystalline defects in lithium-manganese oxide electrodes during electrochemical activation to high voltage, Nat. Commun. 10 (2019) 1–7. https://doi.org/10.1038/s41467-019-09408-2.

[22]  H. Zhang, F. Omenya, P. Yan, L. Luo, M.S. Whittingham, C. Wang, G. Zhou, Rock-Salt Growth-Induced (003) Cracking in a Layered Positive Electrode for Li-Ion Batteries, ACS Energy Lett. 2 (2017) 2607–2615. https://doi.org/10.1021/acsenergylett.7b00907.

[23]   P. Yan, J. Zheng, M. Gu, J. Xiao, J.G. Zhang, C.M. Wang, Intragranular cracking as a critical barrier for high-voltage usage of layer-structured cathode for lithium-ion batteries, Nat. Commun. 8 (2017) 1–9. https://doi.org/10.1038/ncomms14101.

[24]   P. Yan, J. Zheng, T. Chen, L. Luo, Y. Jiang, K. Wang, M. Sui, J.G. Zhang, S. Zhang, C. Wang, Coupling of electrochemically triggered thermal and mechanical effects to aggravate failure in a layered cathode, Nat. Commun. 9 (2018) 1–8. https://doi.org/10.1038/s41467-018-04862-w.

[25]   L. Mu, R. Lin, R. Xu, L. Han, S. Xia, D. Sokaras, J.D. Steiner, T.C. Weng, D. Nordlund, M.M. Doeff, Y. Liu, K. Zhao, H.L. Xin, F. Lin, Oxygen Release Induced Chemomechanical Breakdown of Layered Cathode Materials, Nano Lett. 18 (2018) 3241–3249. https://doi.org/10.1021/acs.nanolett.8b01036.

[26]   U. Ulvestad, A. Singer, J.N. Clark, H.M. Cho, J.W. Kim, R. Harder, J. Maser, Y.S. Meng, O.G. Shpyrko, Topological defect dynamics in operando battery nanoparticles, Science (80-. ). 348 (2015) 1344–1347. https://doi.org/10.1126/science.aaa1313.

[27]   A. Singer, M. Zhang, S. Hy, D. Cela, C. Fang, T.A. Wynn, B. Qiu, Y. Xia, Z. Liu, A. Ulvestad, N. Hua, J. Wingert, H. Liu, M. Sprung, A. V. Zozulya, E. Maxey, R. Harder, Y.S. Meng, O.G. Shpyrko, Nucleation of dislocations and their dynamics in layered oxide cathode materials during battery charging, Nat. Energy. 3 (2018) 641–647. https://doi.org/10.1038/s41560-018-0184-2.

[28]   M. Meyers, K. Chawla, Mechanical Behavior of Materials (2nd ed.), Cambridge University Press, Cambridge, 2008. https://doi.org/10.1017/CBO9780511810947.

[29]   S. Wang, O. Gavalda-Diaz, T. Luo, L. Guo, E. Lovell, N. Wilson, B. Gault, M.P. Ryan, F. Giuliani, The effect of hydrogen on the multiscale mechanical behaviour of a La(Fe,Mn,Si)13-based magnetocaloric material, J. Alloys Compd. 906 (2022) 164274. https://doi.org/10.1016/j.jallcom.2022.164274.

[30]   W.Z. Han, L. Huang, S. Ogata, H. Kimizuka, Z.C. Yang, C. Weinberger, Q.J. Li, B.Y. Liu, X.X. Zhang, J. Li, E. Ma, Z.W. Shan, From "smaller is stronger" to "size-independent strength plateau": Towards measuring the ideal strength of iron, Adv. Mater. 27 (2015) 3385–3390. https://doi.org/10.1002/adma.201500377.

[31]   G. Sernicola, T. Giovannini, P. Patel, J.R. Kermode, D.S. Balint, T. Ben Britton, F.

Giuliani, In situ stable crack growth at the micron scale, Nat. Commun. 8 (2017) 1–9. https://doi.org/10.1038/s41467-017-00139-w.

[32] S. Wang, F. Giuliani, T. Ben Britton, Variable temperature micropillar compression to reveal <a> basal slip properties of Zircaloy-4, Scr. Mater. 162 (2019) 451–455. https://doi.org/10.1016/j.scriptamat.2018.12.014.

[33] S. Wang, F. Giuliani, T. Ben Britton, Slip–hydride interactions in Zircaloy-4: Multiscale mechanical testing and characterisation, Acta Mater. 200 (2020) 537–550. https://doi.org/10.1016/j.actamat.2020.09.038.

[34] M.D. Uchic, D.M. Dimiduk, J.N. Florando, W.D. Nix, Sample dimensions influence strength and crystal plasticity, Science (80-. ). 305 (2004) 986–989. https://doi.org/10.1126/science.1098993.

[35] W. Huddleston, F. Dynys, A. Sehirlioglu, Effects of microstructure on fracture strength and conductivity of sintered NMC333, J. Am. Ceram. Soc. 103 (2020) 1527–1535. https://doi.org/10.1111/jace.16829.

[36] T. Yoon, M.S. Milien, B.S. Parimalam, B.L. Lucht, Thermal Decomposition of the Solid Electrolyte Interphase (SEI) on Silicon Electrodes for Lithium Ion Batteries, Chem. Mater. 29 (2017) 3237–3245. https://doi.org/10.1021/acs.chemmater.7b00454.

[37] L.N. Dinh, W. McLean, M.A. Schildbach, J.D. LeMay, W.J. Siekhaus, M. Balooch, The nature and effects of the thermal stability of lithium hydroxide, J. Nucl. Mater. 317 (2003) 175–188. https://doi.org/10.1016/S0022-3115(03)00084-9.

[38] T. Roisnel, J. Rodríquez-Carvajal, WinPLOTR: A Windows Tool for Powder Diffraction Pattern Analysis, Mater. Sci. Forum. 378–381 (2001) 118–123. https://doi.org/https://doi.org/10.4028/www.scientific.net/MSF.378-381.118.

[39] W.C. Lenthe, S. Singh, M. De Graef, A spherical harmonic transform approach to the indexing of electron back-scattered diffraction patterns, Ultramicroscopy. 207 (2019) 112841. https://doi.org/10.1016/j.ultramic.2019.112841.

[40] M.B. Lowry, D. Kiener, M.M. Leblanc, C. Chisholm, J.N. Florando, J.W. Morris, A.M. Minor, Achieving the ideal strength in annealed molybdenum nanopillars, Acta Mater. 58 (2010) 5160–5167. https://doi.org/10.1016/j.actamat.2010.05.052.

[41] S. Shim, H. Bei, M.K. Miller, G.M. Pharr, E.P. George, Effects of focused ion beam

milling on the compressive behavior of directionally solidified micropillars and the nanoindentation response of an electropolished surface, Acta Mater. 57 (2009) 503–510. https://doi.org/10.1016/j.actamat.2008.09.033.

[42] D. Kiener, C. Motz, M. Rester, M. Jenko, G. Dehm, FIB damage of Cu and possible consequences for miniaturized mechanical tests, Mater. Sci. Eng. A. 459 (2007) 262–272. https://doi.org/10.1016/j.msea.2007.01.046.

[43] K. Thompson, D. Lawrence, D.J. Larson, J.D. Olson, T.F. Kelly, B. Gorman, In situ site-specific specimen preparation for atom probe tomography, Ultramicroscopy. 107 (2007) 131–139. https://doi.org/10.1016/j.ultramic.2006.06.008.

[44] J.P. McCaffrey, M.W. Phaneuf, L.D. Madsen, Surface damage formation during ion-beam thinning of samples for transmission electron microscopy, Ultramicroscopy. 87 (2001) 97–104. https://doi.org/10.1016/S0304-3991(00)00096-6.

[45] S. Lee, J. Jeong, Y. Kim, S.M. Han, D. Kiener, S.H. Oh, FIB-induced dislocations in Al submicron pillars: Annihilation by thermal annealing and effects on deformation behavior, Acta Mater. 110 (2016). https://doi.org/10.1016/j.actamat.2016.03.017.

[46] H. Bei, S. Shim, M.K. Miller, G.M. Pharr, E.P. George, Effects of focused ion beam milling on the nanomechanical behavior of a molybdenum-alloy single crystal, Appl. Phys. Lett. 91 (2007) 1–4. https://doi.org/10.1063/1.2784948.

[47] M. Higashi, S. Momono, K. Kishida, N.L. Okamoto, H. Inui, Anisotropic plastic deformation of single crystals of the MAX phase compound Ti3SiC2 investigated by micropillar compression, Acta Mater. 161 (2018) 161–170. https://doi.org/10.1016/j.actamat.2018.09.024.

[48] O. Gavalda-Diaz, J. Lyons, S. Wang, M. Emmanuel, K. Marquardt, E. Saiz, F. Giuliani, Basal Plane Delamination Energy Measurement in a Ti3SiC2 MAX Phase, Jom. 73 (2021) 1582–1588. https://doi.org/10.1007/s11837-021-04635-9.

[49] F.P.E. Dunne, N. Petrinic, Introduction to Computational Plasticity, Oxford University Press, Oxford, 2005.

[50] F. Östlund, K. Rzepiejewska-Malyska, K. Leifer, L.M. Hale, Y. Tang, R. Ballarini, W.W. Gerberich, J. Michler, Brittle-to-ductile transition in uniaxial compression of silicon pillars at room temperature, Adv. Funct. Mater. 19 (2009) 2439–2444.

https://doi.org/10.1002/adfm.200900418.

[51] P.R. Howie, S. Korte, W.J. Clegg, Fracture modes in micropillar compression of brittle crystals, J. Mater. Res. 27 (2012) 141–151. https://doi.org/10.1557/jmr.2011.256.

[52] O. Gavalda-Diaz, M. Conroy, F. Giuliani, Imaging Stress Induced Domain Movement and Crack Propagation by in situ Loading in the Transmission Electron Microscope, Microsc. Microanal. 28 (2022) 2340–2340. https://doi.org/10.1017/s1431927622008984.

[53] M.D. Radin, S. Hy, M. Sina, C. Fang, H. Liu, J. Vinckeviciute, M. Zhang, M.S. Whittingham, Y.S. Meng, A. Van der Ven, Narrowing the Gap between Theoretical and Practical Capacities in Li-Ion Layered Oxide Cathode Materials, Adv. Energy Mater. 7 (2017) 1–33. https://doi.org/10.1002/aenm.201602888.

# Supplementary tables and figures

Table S1 Le Bail refined lattice parameters and primitive cell volume of sintered NMC811. The $\chi^2$ value indicates good fit of experimental data to calculated pattern.

| a (Å) | c (Å) | Cell volume (Å$^3$) | $\chi^2$ |
|---|---|---|---|
| 2.8954(1) | 14.2749(6) | 103.64(1) | 2.20 |

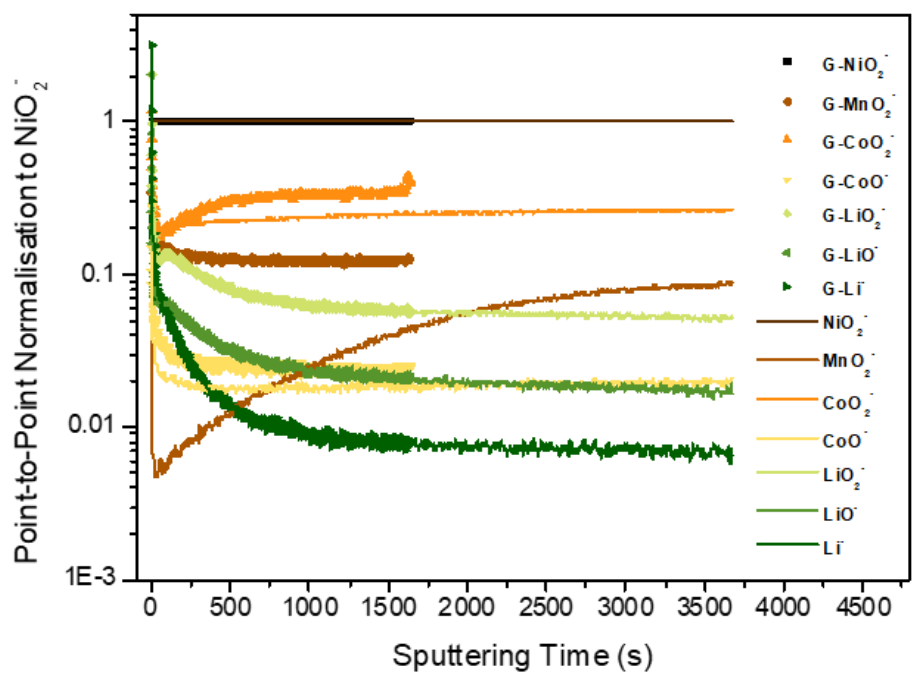

Figure S1  SIMS depth profiles for the NMC811 pellets before and after sintering. The symbols present the data collected on the green pellets before sintering, while the solid lines present the data obtained on the pellets after sintering at 1000 °C. The two datasets were collected under the same experimental measurement conditions and are point-to-point normalised to $NiO_2^-$ signals. For Li-related species ($LiO_2^-$, $LiO^-$ and $Li^-$) the depth profiles before and after sintering agree well with each other, therefore compared to Ni-related species no obvious change in Li-related species before and after sintering is observed.

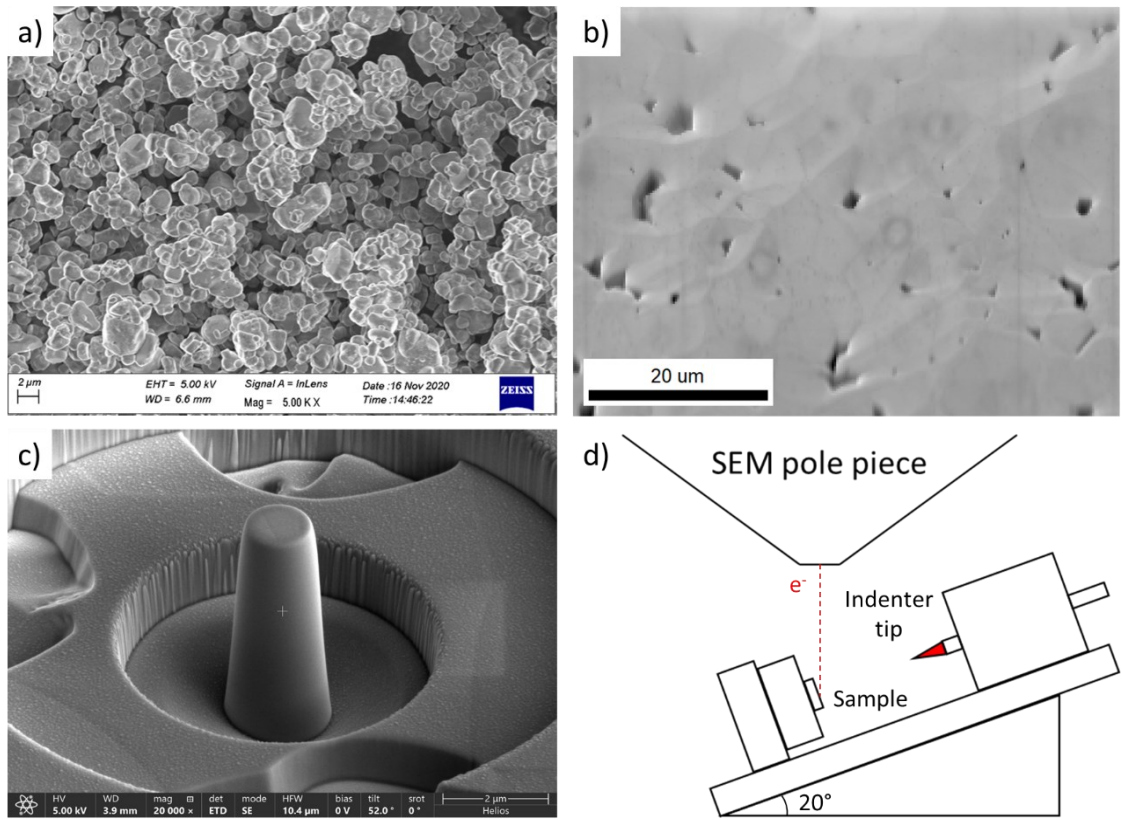

Figure S2 SEM images of (a) single crystal NMC811 powders used for sintering, (b) a sintered pellet (corresponding to the IPF map in Figure 1) and (c) an uncompressed single crystal NMC 811 micropillar. (d) Schematic diagram of the in situ SEM micropillar compression experimental setup.

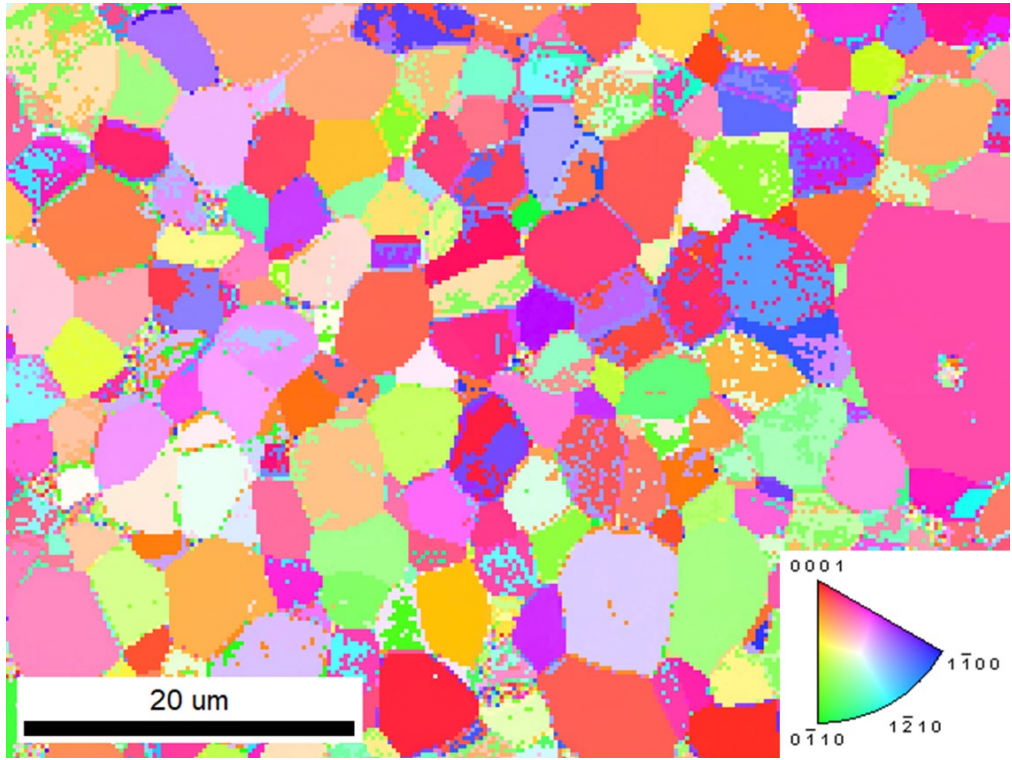

Figure S3  EBSD IPF-z map (z points out of the page / sample surface) of sintered NMC811, plotted using the raw data.

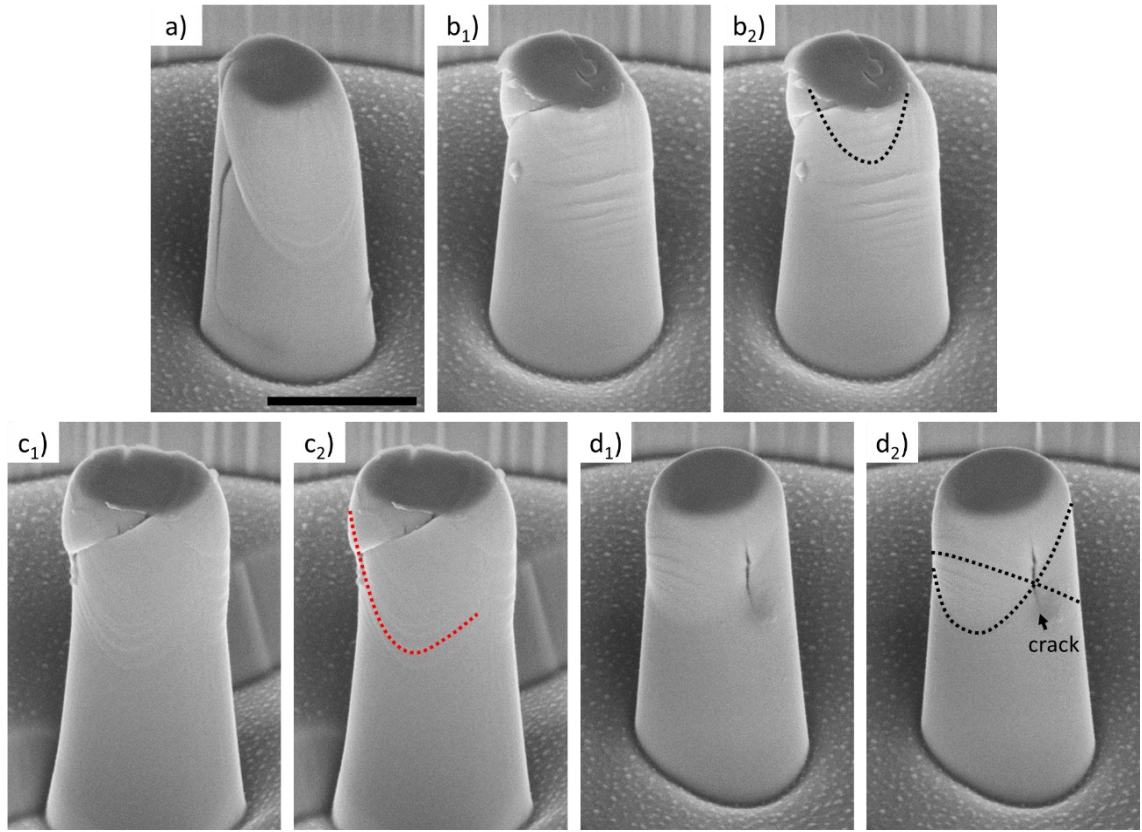

Figure S4 SEM images captured from alternative viewing angles for micropillars in (a) Figure 2(a), (b) Figure 2(c) and (c) Figure 3(b). (d) SEM image of a tested single crystal NMC811 pillar showing a vertical crack at intersections between two sets of pyramidal slip bands. The slip traces (mentioned in the text) on the pillars in ($b_1$), ($c_1$) and ($d_1$) are marked with dotted lines in ($b_2$), ($c_2$) and ($d_2$), respectively. Basal and pyramidal slip traces are marked in red and black, respectively. The scale bar is 1 μm for all images.

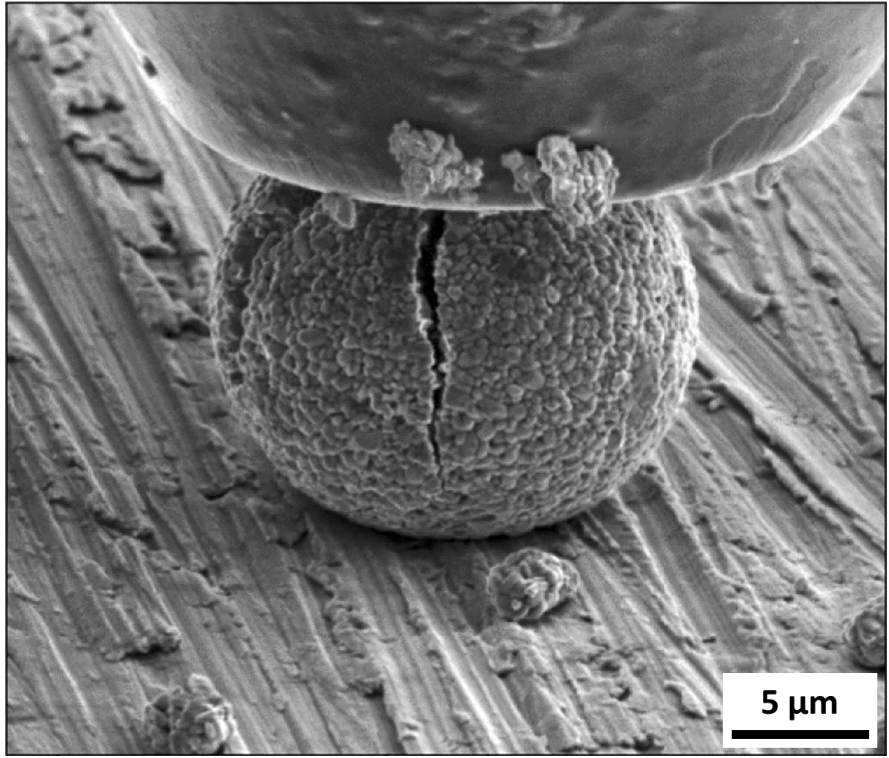

Figure S5 Real-time SEM image recorded during a flat punch compression test of an agglomerated polycrystalline NMC811 particle showing vertical cracks.

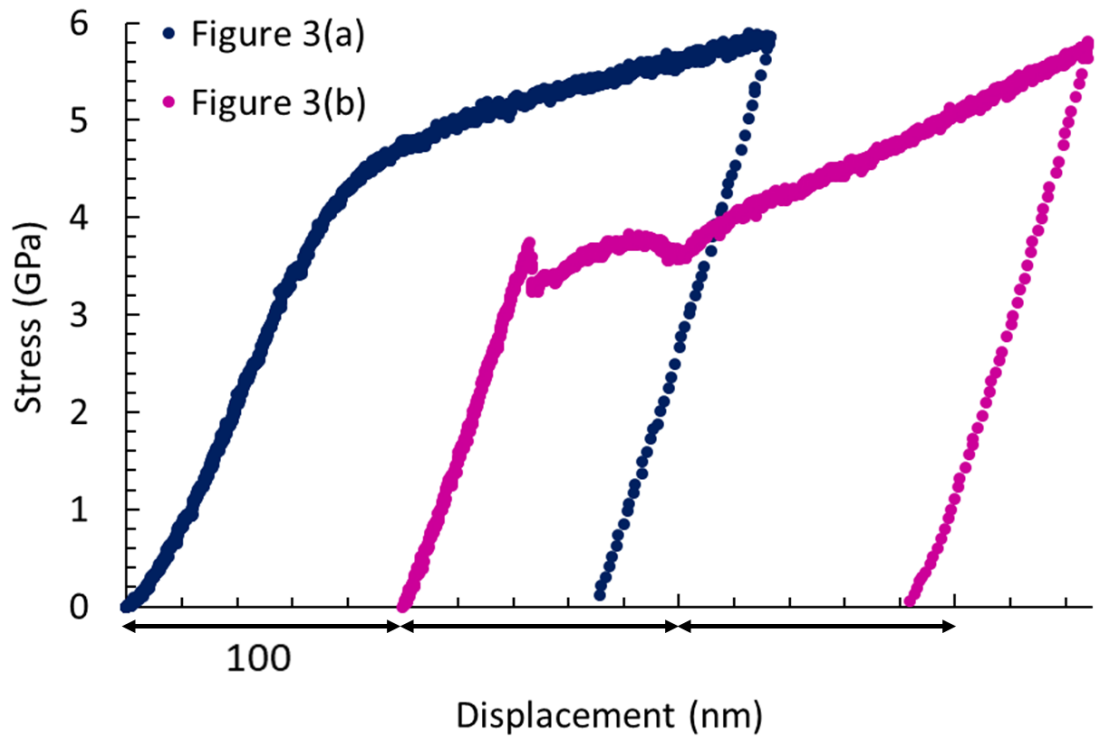

Figure S6 Stress-displacement curves recorded during compression tests of the single crystal NMC811 micropillars displayed in Figure 3.

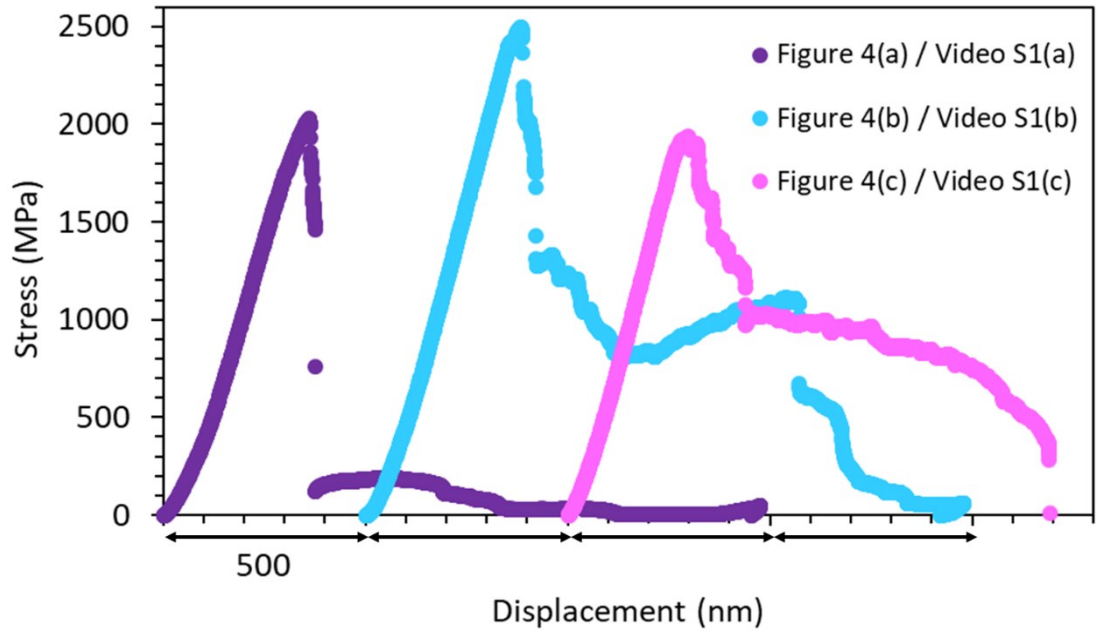

Figure S7  Stress-displacement curves recorded during compression tests of the polycrystalline NMC811 micropillars displayed in Figure 4 and Video S1. For all three pillars, stress drops occurred immediately after the elastic regimes and no stable plastic flow can be observed.